\documentclass{aastex631}

\begin{document}

\title{Party Planning the Next True Happy New Year: Lunar Orbital Evolution Epochs with Integer Synodic Months Per Year}

\shorttitle{\sc{True Happy New Year Party Planning}}
\date{April 1st 2023}

\correspondingauthor{Mark Popinchalk}
\author[0000-0001-9482-7794]{Mark Popinchalk}
\affiliation{\amnh}
\affiliation{\cunygrad}
\affiliation{\hunter}
\email{popinchalkmark@gmail.com}

\newcommand{\amnh}{Department of Astrophysics, American Museum of Natural History, Central Park West at 79th Street, New York, NY 10034, USA}
\newcommand{\cunygrad}{Physics, The Graduate Center, City University of New York, New York, NY 10016, USA}
\newcommand{\hunter}{Department of Physics and Astronomy, Hunter College, City University of New York, 695 Park Avenue, New York, NY 10065, USA}

\begin{abstract}
Humans like to party, and New Year celebrations are a great way to do that. However New Years celebrations that rely on an orbital year don't line up with those that use a Lunar Calendar, as there are currently 12.368 synodic months (moonths) in a year. There is cyclostratigraphic, paleontological, and tidal rhythmite data that reveal that over billions of years the interplay of angular momentum between the Sun, Earth and Moon has changed the rate of rotation of Earth, and at the same time evolved the orbit of the Moon, and therefore the length of a Lunar month. Using a subset of this data and referencing literature models of the Moon's orbital evolution, we create our own simple model to determine "True Happy New Years", time periods when there were an integer number of lunar synodic months in an Earth orbital year. This would allow modern calendars to pick a shared New Year's Day, and party accordingly. We then predict the next True Happy New Year to be in 252 million years, and offer suggestions to begin the party planning process early, so that we as a planet may be ready.
\end{abstract}

\section{Introduction}

Göbekli Tepe is a Neolithic site in Turkey, and a location where humans were thought to have celebrations over 10,000 years ago, before the domestication of animals or even agriculture \cite{gobekli}. It must be assumed without reference however, that humans have been looking for an excuse to have a little treat, festival, or feast for at least 10,000 years before that, if not 100,000 years. It is a fundamental aspect of human life that "we like to party" (Vengaboys, priv. comm).

Many of these parties would have had to do with the passage of time, celebrating seasonal or annual events. Astronomy played a role in early calendars, as recognizing when certain constellations rose and set in the sky could be associated with certain seasons, and a lunar orbit is a convenient amount of time and a significant fraction of the year. Indeed, the English word for month carries its lunar history and basis (moonth).

In our modern world, it is no different. As a species we still like to party, and many modern festival, feasts, and fetes are based around the completion of one "year". However, they are not generally unsynchronized, and happend at different times throughout the year. One major celebration is at the end of a "calendar year", which attempts to match the Earth's orbital period. These "New Years Days" are celebrated every 365 days, with the need for as many as 20 intercalary days added over the course of an average lifetime to match the real orbital period. 

This contrasts with a "lunar new year", which is celebrated either on the first instance that a lunar phase lines up within a new calendar year, requiring intrecalary weeks each year. Other "lunar new years" are defined as a distinct number of lunar synodic months (here after referred to as moonths), for example 12 moonths in a year. This will necessarily move their start date 10-12 days over each Earth orbit, as there are current 12.368 moonths per year.

These differences are in part due to the fact that the number of days in a year do not add up to an integer number of moonths. At least not currently. The Earth-Moon-Sun system has served to shuffle angular momentum throughout the three bodies, which has ultimately led to the length of the Earth's day and the orbit of the Moon to change over billions of years. In this paper, we will examine when in Earth-Moon systems past was there the opportunity to have a "True Happy New Year", where an integer number of moonths precisely equals the time of an Earth orbit or year. Most importantly we will predict the 12 moonth True Happy New Year, which will require exceptional planning \citep[see][]{party}{}{}

In Section~\ref{sec:evolution} we detail our efforts to describe the evolution of the Earth-Moon System, Section~\ref{sec:results} we present our calculations for the previous 14 and 13 moonth True Happy New Years as well as the upcoming 12 moonth True Happy New Year and Section~\ref{sec:discussion} we will describe steps we should take to plan for it.

\section{Evolution of the Earth-Moon System}\label{sec:evolution}

\subsection{Definition of a Year}

The current orbital period of the Earth-Moon barycenter is 365.256 days, and conveniently orbits at precisely 1 AU. \cite{inso_2004} calculates that this orbital distance does vary, and ranges atmost from .99998 to 1.00002, representing a $\sim$ 7 minute change in the duration of year, which we will consider negligible and assume a fixed value for an Earth year of 365.256 days.

It is important to note that this is a sidereal year, defined against an astronomical frame of reference (comparing Earths position to distant stars). A tropical year defined from Earth-Sky coordinates such as the equinoxes is 365.242 days (as axial precession “moves” the sky slightly over a year), and an anomalistic year that uses an Earth-Sun reference is 365.260 (apsidal precession “moves” perehelion).

Our simple model will not take into account the eccentricity, axial tilt or precession of the Earth’s orbit's, forcing it to be fixed. Furthermore, we will not be considering any other factors, such as they are beyond the scope of this work. In effect, we assume the Earth’s year is fixed at a single number, 365.256 days.

\subsection{Data}

Evidence for the evolution of the Earth-Moon system comes from many sources. We present the data we compiled for our analysis in terms of the three methods or field from which it comes, and describe briefly the method involved

\subsubsection{Cyclostratigraphic}

While we made the assumption that the Earth's orbital parameters are fixed in the previous section, that is not the case. Gravitational interactions between the Earth and the Sun, other planets, and most importantly the Moon alter it's eccentricity, axial precession rate, and obliquity. These processes are cyclical, and are dubbed Milankovitch Cycles \cite{miloankovith} and appear in coral reefs and deep sea sediments \citep{1968Sci...159..297B}

Milankovitch cycles also change the amount of solar insolation on the surface of the planet, which in turn will effect the rate and formation of sedimentary layers in geologic strata. Cyclostratigraphy is a method to measure the Milankovitch cycles through strata on the Earth. From an understanding of the Milankovitch cycle, the raw measurement is the rate of precession of the Earth, from which a Lunar distance can be calculated. This method is an exciting new prospect for dating historical climates. \citep[][]{cyclo_fun}{}{}

We use the data compiled in \cite{ocean_main}, namely from \cite{cyclo_1,cyclo_2,cyclo_3,cyclo_4,cyclo_5}. We convert the calculated Lunar semi-major axes into moonths using Kepler's Laws.

\subsubsection{Paleontology}

Animals grow over time, and often with annual and lunar cycles. Mollusk shells will daily, and be sensitive to the higher tides that come with the Lunar cycle. It is therefore possible to count the number of days in a lunar month, as well as the number of moonths per year. \cite{shells}. This is also true for stromatolites \cite{stromato} and other animals.

We use the data in \cite{palentology_1,palentology_2}, compiled in Wu et al. (submitted).

\subsubsection{Tidal Rhythmites}

Finally, deposits of material on the edge of a body of water will be sensitive to annular, lunar, and daily forces. For particular bodies of water, the deposit of material from the water onto surrounding rock will be determined by the water levels and therefore tidal activity. It is therefore possible to examine individual deposits to measure the lunar cycle, and a value for moonths per year.

We use the data in \cite{tidalrhytmites_1,tidalrhytmites_2,tidalrhtymites_3}, drawn from Wu et al. (submitted) and \cite{ocean_main}.

\subsection{Modeling the evolution}

\begin{figure}
\begin{center}
\includegraphics[width=0.85\linewidth]{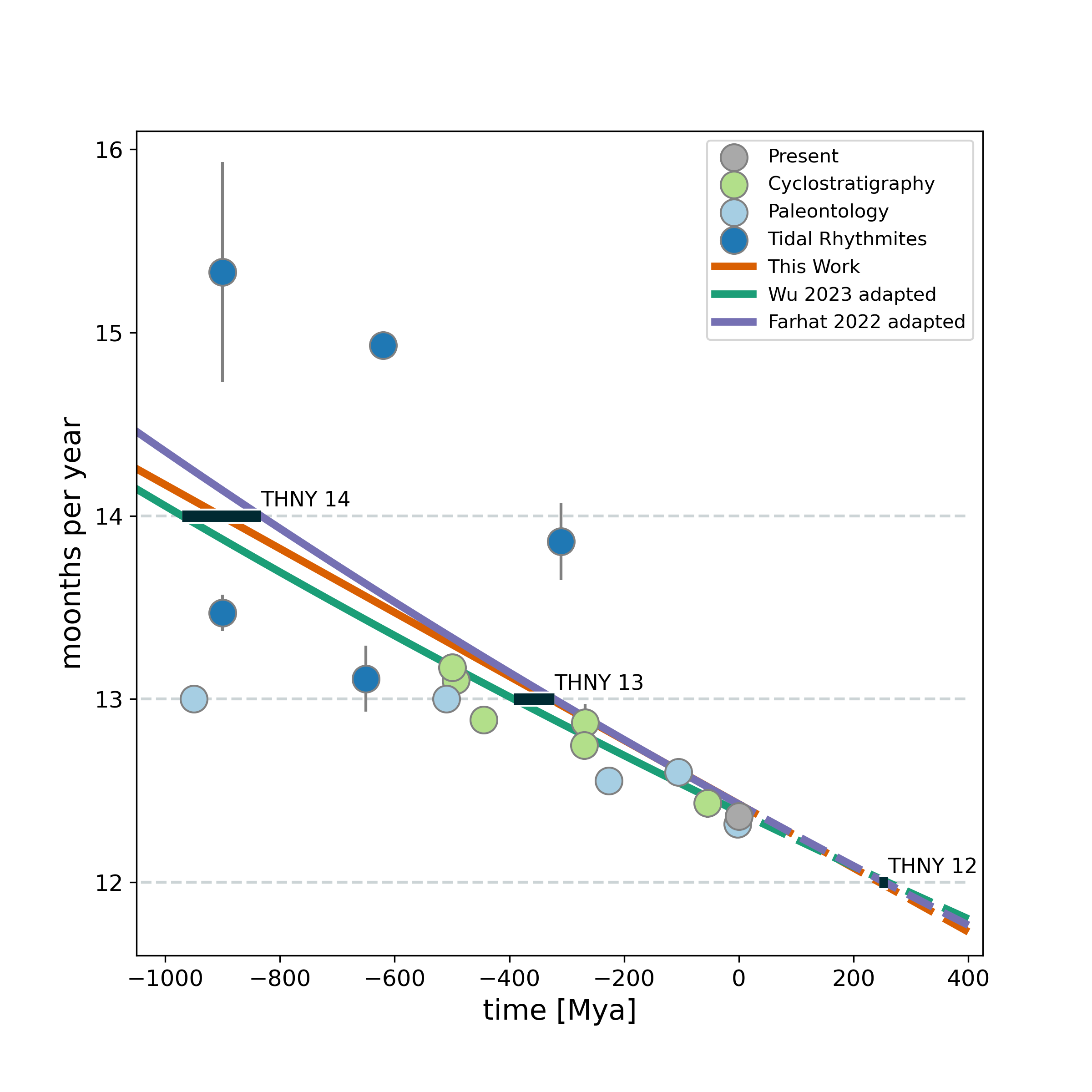}
\caption{Measurements and models for the number of moonths per year over the last billion years and in to the future. Data points are color coded by method. Our linear model is presented in orange and conversions of the Wu et al. (submitted) and \cite{ocean_main} are green and blue respectively. Dashed lines matching their colors carry the models into the future. Black lines with white outlines are labeled with THNY for True Happy New Years where models cross the integer moonth per year sections of the plot.}
\label{fig:tny}
\end{center}
\end{figure}

In this work, we only consider data that is dated to the last billion years. Wu et al. (submitted) comment that beyond that date the data become challenging to use, and for our party planning purposes we focus on time periods where there might be at least multi-cellular life. We present this data in \ref{fig:tny}.

We fit a line to the number of moonths per year over time. This is a dramatic oversimplification of the rich and interesting history of the angular momentum evolution of the Earth-Moon system. If interested in learning more, we suggest reading \cite{ocean_main}, which model the changing surfaces of the planet and therefore the tidal effects of different ocean arrangements and how that effects the evolution. We also suggest Wu et al. (submitted, presented at AAS 241 by N. Murray 2023) which not only the torque from the Moon, but also from the Sun due to thermal tides, which may imply a resonant length of day at certain points in the Earth's history.

From Wu et al. (submitted) and \cite{ocean_main} we make a crude linear assumption of their Lunar semi-major axis calculations for the last billion years, and convert them to moonths per year, to compare against our own.

\section{Results - The True Happy New Years}\label{sec:results}

The three models used to calculate the True New Years (by eye conversions of Wu et al. (submitted) and \cite{ocean_main} and our own linear model) each report an age for when the True Happy New Years would have occurred. For the sake of simplicity, we report an average age of -902.343 $\pm$ 56.002 Mya for the 14 moonth True Happy New Year, and -347.964 $\pm$ 31.601 Mya for the 13 moonth True Happy New Year.

This puts the 14 moonth True Happy New Year in the Tonian period of the Proterozoic era. Multi-cellular life was developing, and there may be evidence of early sponges existing during this time period \cite{sponge}. It is however in a time period dubbed "The Boring Billion", which does not bode well for the party.

The 13 moonth True Happy New Year would have then likely have occurred in the Carboniferous period during the Paleozoic era, in particular the Visian age. This is a geologic period where tetrapod fossils are not as widely available known as Romer's Gap \citep{romers}, but there is evidence of reptile like amphibians living at that time \citep{weskothina}.

Finally, by propagating the models forward we can estimate when the 12 moonth True New Year will occur. We note that this was not the intention of Wu et al. (submitted) and \cite{ocean_main} models, which were constructed to describe the past evolution of the Earth Moon system. However \cite{inso_2004} do predict the next 250 Mya of Earth Moon evolution, and it appears consistent with at least the last 250 Mya. We therefore report an age of 252.404 Mya $\pm$ 6.029 for the 12 moonth True Happy New Year for the Earth Moon System.

\section{Discussion - Party Planning the 12 moonth True Happy New Year}\label{sec:discussion}
\subsection{What Makes a Good Party?}

While we have discussed the opportunity for a party that the True New Year offers for those who are able to witness it, we have yet to describe what makes a great party.

Fortunately we can look at the previous True New Years for direction. We begin by positing that Liveliness of a party is important. The 14 moonth True New Year was populated by simple multicellular life, which don't have mouths that can speak and likely would have made interesting conversation challenging. The 13 moonth True New Year did have a wider range of animal species present, including trilobites, fishes, and early tetrapods. This would have been a vast improvement in terms of entertainment, but would have kept celebrations in the ocean, or at least near the water's edge. This means that it was ultimately still be a pool party, which can be polarizing and brings with it the question are people responsible for their own towels\citep[see][]{douglas_adams}. Animal species have radiated into fascinating forms in the hundreds of millions of years since. Despite the best efforts of our anthropocentric vice of decreasing biodiversity across the world for the last 10000+ years, there is good reason to believe that a range of lively animals will exist as potential guests at the time of the 12 moonth True New Year.

Furthermore, a second fundamental aspect of a party is that it be Interesting. During the 12 moonth True New Year, not only will New Years Day fall on the same Lunar phase every year, but so will one's birthday. While surely an opportunity for pseudoscientific grabble to proliferate, each person would have a Lunar phase associated to them throughout their lives. It could enable a kind of fragmented celebration throughout the year - instead of forcing a birthday celebration into a day or week, one would be able to point to fractional. Interesting!

Finally, Timeliness is incredibly important for a good party experience. While start time for social invitations can be difficult to judge (there is still no scientific definition for Party-o-clock), it is far easier to notice when a party has gone on for too long, or ends too soon. This is a subjective measurement, and will change with observer. Fortunately, the duration of the 12 moonth True New Year will be ~50,000 years longer than the 13 moonth True year due to the rate of recession decreasing, which if we assume an average lifespan of 80 years, means that an additional 625 generations of humans can struggle with the decision of when the right time is to leave.

And so, with this combination of Liveliness, Interest, and Timeliness, we speculate that the 12 moonth True New Year will be thoroughly LIT.

\subsection{In the Meantime}

It is not our time to enjoy a True Happy New Year. Some may feel that with our civilization's current trajectory, it seems far, distant, and faint. Planning for a party that one will never see might seem foolish, but it implies an optimistic outlook that our species can be a long lived one, capable of coexisting on this planet for astronomical timescales, orders of magnitude longer than it has existed so far. I hope that when the next True Happy New Year comes, there is someone to mark the significance.

\section{Conclusions}

We have coined the idea of a "True Happy New Year", when the Earth-Moon system are in an arrangement such that there are an integer number of synodic Lunar months in an Earth Year. We use a simple analysis of cyclostratigraphic, paleontological, and tidal rhythmite data to assume a linear change in the number of moonths per year over the last billion years. This agrees somewhat with more refined models, which we compare against by assuming their models are linear over that time frame. Using these model we describe the party guests of the 14 and 13 moonth True New Year, sponges (boring) and early amphibians (wet) respectively. Finally, we predict the 12 moonth True New Year to occur in 252 million years. This would allow many modern calendars systems to converge, which would be an opportunity for a big party, which we predict to be LIT. 

\begin{acknowledgments}
The author would like to thank Professor N. Murray for sending an early version of Wu et al. and the accompanying data.
\end{acknowledgments}

\bibliography{bilbo}{}
\bibliographystyle{aasjournal}

\end{document}